\documentclass[conference]{IEEEtran}
\usepackage{graphicx}

\usepackage{subfigure} % Written by Steven Douglas Cochran
\begin{document}

\title{Detection and measurement of gamma rays with the AMS-02
detector}

\author{\authorblockN{ Simonetta Gentile}
\authorblockA{Dipartimento di Fisica,  Universit\`a La Sapienza, Sez. I.N.F.N., Roma,\\
Piazza A.Moro 2,
00183, Roma (Italy)\\
Email: simonetta.gentile@roma1.infn.it\\
\\
on behalf of AMS-02 collaboration
}}

% avoiding spaces at the end of the author lines is not a problem with
% conference papers because we don't use \thanks or \IEEEmembership

%\specialpapernotice{(Invited Paper)}

\maketitle

\begin{abstract}
The Alpha Magnetic Spectrometer (AMS-02)  will be installed on the
International Space Station (ISS).
The gamma rays can be measured through gamma conversion into ${\rm e^+e^-}$ pair, before
reaching the Silicon Tracker or by measurement of a photon hitting directly
the Electromagnetic Calorimeter (ECAL).
AMS-02 will provide precise gamma measurements in the GeV energy range,
which is particularly relevant for Dark Matter searches.
In addition, the good angular resolution and identification capabilities of the detector
will allow studies of the main galactic and extra-galactic sources, diffuse gamma
background and Gamma Ray Bursts.

\end{abstract}


% for peerreview papers, inserts a page break and creates the second title.
% Will be ignored for other modes.
\IEEEpeerreviewmaketitle

\section{Introduction}
% no \PARstart

\par The AMS-02  detector \cite{Barao:2004ik}
is  optimised for detection of charged cosmic rays and it is able to identify gamma rays.
It is sensitive to photons in the  energy range from 1 GeV up to a few hundred GeV poorly 
covered by other experimental data. Its large acceptance will allow to fill 
this gap.
%with a large ammount of
% experimental data  this energy range.
\par  The most  precise data  available in the GeV energy range  at present are EGRET measurements of 
gamma ray fluxes performed in 1990s'  \cite{EGRET,Hunger:1997we,3eg}. 
These date have motivated a large number of studies. 
%and reinterpretation. 
To prove the validity of the various  gamma emission models more precise measurements are needed.  
The understanding of emissions from gamma sources (pulsars, blazars, AGNs),
diffuse gamma background emission, Dark Matter searches
and Gamma Ray Bursts (GRB) demand more accurate data.

%MS: the line below should go somewhere else! 
%\hfill November 1, 2006
\section {Detector performances}
\subsection{AMS-02 detector}\label{sec:AMS}
\begin{figure}[t]
  \begin{center}
\includegraphics*[width=0.4\textwidth]{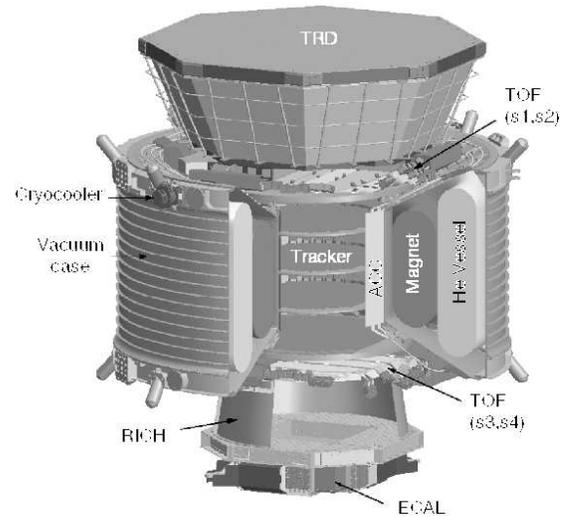}

  \end{center}
  \vspace{-0.5pc}
  \caption{The Alpha Magnetic Spectrometer. The detector components are: Transition Radiation Detector(TRD),
 Time-of-flight Scintillators(TOF), Silicon Tracker (Tracker), Ring Imaging Cerenkov detector(RICH),
lead/plastic fiber calorimeter(ECAL), 
the anticoincidence counters(ACC) are located in inner side in the magnet.
}
\label{fig:AMS}
\end{figure}

\par Fig. \ref{fig:AMS} shows the
     AMS-02 detector. The  main components are:

\begin{itemize}
  \item   A 20 layer Transition Radiation Detector (TRD) to distinguish protons/antiprotons from positrons/electrons
 with a rejection factor of $10^2-10^3$ in a range from 1.5 to 300 GeV.
 The TRD will be used in conjunction with the electromagnetic calorimeter to provide an overall $\mathrm {e^+}$/p rejection $10^{-6}$.

 \item Four layers of Time of Flight (TOF) hodoscopes  provide precision time of flight
     measurements ($\sim$ 120 ps), d$E$/d$x$ measurements and  trigger.

  \item   The superconducting magnet which provides a bending power of $\mathrm{BL^2} \sim 0.8\mathrm { T m^2}$.

  \item   Eight layers (6.45 $\mathrm m^2$) of double-sided silicon tracker to provide a coordinate %MS : better 'spacial'
     resolution of 10$\rm \mu m$ in the bending plane and 30 $\rm \mu m$ in the non-bending plane and d$E$/d$x$ measurements.

  \item   Veto counters to ensure that only particles passing the magnet aperture and not being scattered in the tracker 
will be accepted.

  \item   A Ring Imaging Cerenkov Counter (RICH) which measures the velocity to 0.1 $\%$
     accuracy of particles or nuclei and $\vert Q \vert$. This information combined with the momentum measurement
      in the magnet, will enable AMS-02 to directly measure the mass of particles
     and nuclei and to discriminate isotopes.

  \item   A 3-D sampling calorimeter (ECAL) made out of $16X_0$ of lead and plastic fibbers to
     measure the energy of gamma rays, electrons and positrons and to distinguish electrons
     and positrons from hadrons with a rejection of $10^3$ or $10^4$, combining the tracker information, in the range 1.5 GeV-1 TeV.
\end{itemize}
\par Thus the value of the particle charge $\vert Q \vert$ is measured independently in the Tracker, RICH
     and TOF. The signed charge ($\pm Q$) and the momentum of the particle are measured by  silicon tracker in the magnet. The velocity, $\beta$, is measured by the
     TOF and RICH. Hadron rejection is provided by TRD, ECAL and tracker. 

 The detector is designed with the following properties:
%\begin{itemize}
%   \item 
   minimal material in the particle trajectory so that the material itself is not a source of
     background nor a source of large angle nuclear scattering;
    % \item
 many repeated measurements of momentum and velocity so as to ensure that particles
     which experience large angle nuclear scattering within the detector be swept away by the
     spectrometer and not confused with the signal;
  %\item 
  a geometrical acceptance %in  solid angle 
  of 0.5 $\mathrm{ m^2 }$ sr for the $\mathrm{\overline He}$  search;
%\item 
     hadron/positron rejection better than $\rm 10^6$;
$\rm \Delta \beta$/ $\beta$ =0.1$\%$ to distinguish $^9$Be, $^{10}$Be, and $^3$He, $^4$He light isotopes;
%\item  
   a proton rigidity, $\rm R=pc/ \vert Z\vert e$ (GV), resolution of 20$\%$ at 0.5 TV and
 a helium resolution of 20$\%$ at 1 TV.

These characteristics allows high precision measurements of charged
particle  momenta and masses and energy and angles of photons.
%\end{itemize}

\subsection{Performance of photon detection}\label{subsec:phot_det}
\begin{figure}[h]
%\begin{figure*}
\centerline{\subfigure[{\it Conversion mode}]{\includegraphics[width=0.22\textwidth]{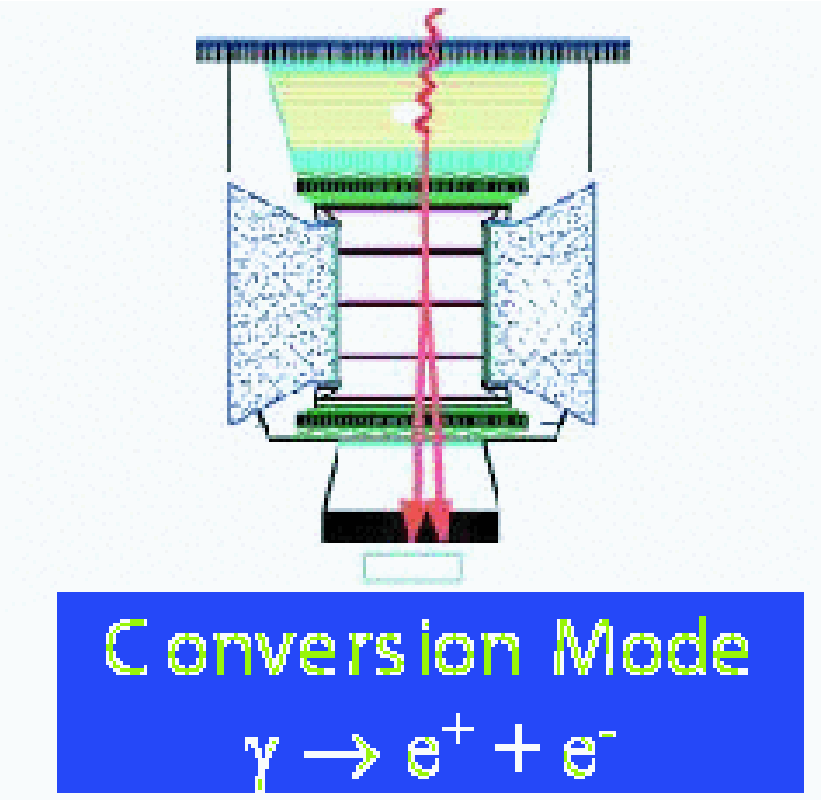}
\label{fig:gamma_tracker}}
\hfil
\subfigure[{\it Calorimetric mode}]{\includegraphics[width=0.22\textwidth]{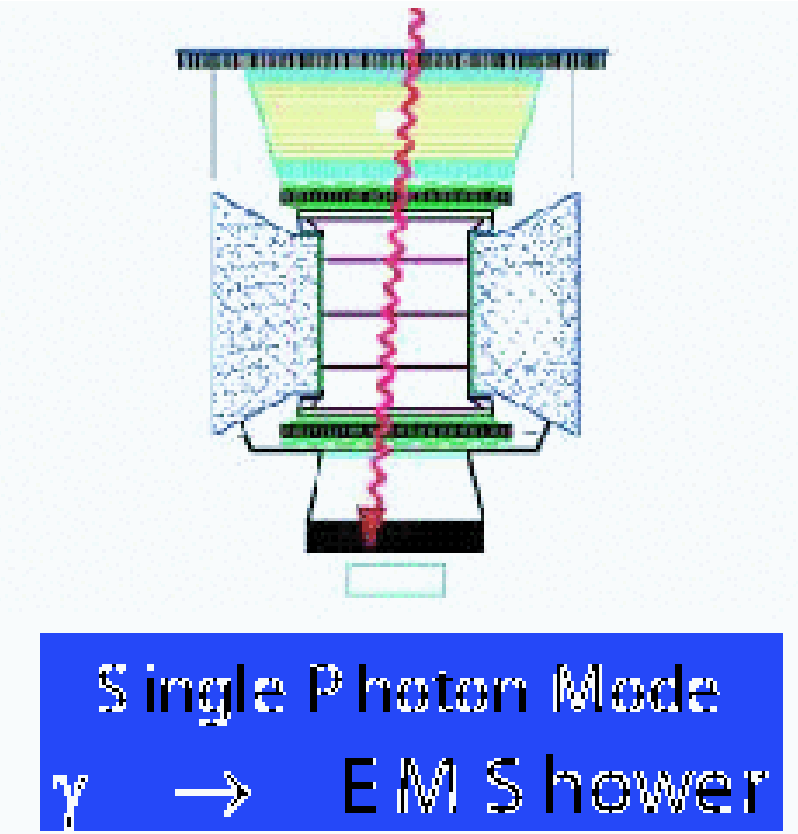}
\label{fig:gammaecal}}}
\caption{ The two modes of gamma detection in AMS-02.}
\label{fig_mode}
%\end{figure*}
\end{figure}
AMS-02 is able to detect gamma rays in two complementary modes:
\begin{itemize}
\item  {\it Conversion mode}. Photons are converted
into electron-positron pairs in the Transition Radiation Detector (TRD) material
(about $0.25X_0$), before reaching the first Time-of-Flight (TOF) layer.
The electron-positron pair is triggered by the TOF system. The tracks of electron and positron
are reconstructed by the tracker. In this mode the viewing angle is about 42 degrees, \hbox {Fig. \ref{fig:gamma_tracker}}.

\item  {\it Calorimetric mode}. Photon which passes through \hbox{AMS-02} without interaction are measured in
the electromagnetic calorimeter.
A special trigger  based on this calorimeter has been developed.
The viewing angle in this, so called, "ECAL mode" is about 23 degrees \hbox {Fig. \ref{fig:gammaecal}}, .
\end{itemize}

\begin{figure*}
\centerline{\subfigure[ Acceptance as a function of energy]{\includegraphics[width=0.34\textwidth]{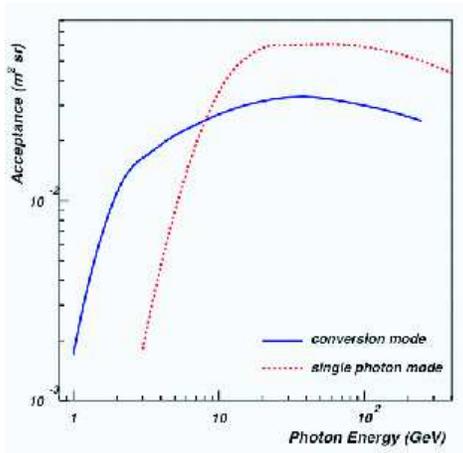}
\label{fig:Acceptance}}
\hfil
\subfigure[ Effective area as a function of cosinus of incident angle]{\includegraphics[width=0.34\textwidth]{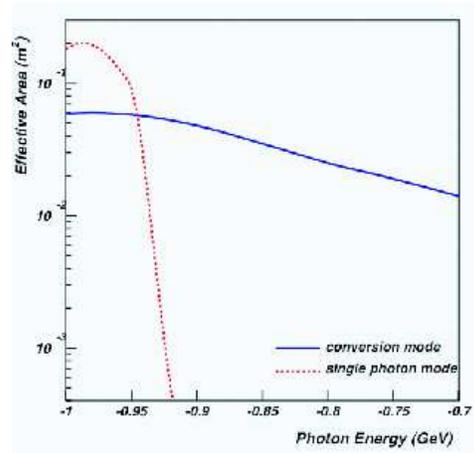}
\label{fig:Effective}}}

\centerline{\subfigure[ Energy resolution]{\includegraphics[width=0.44\textwidth]{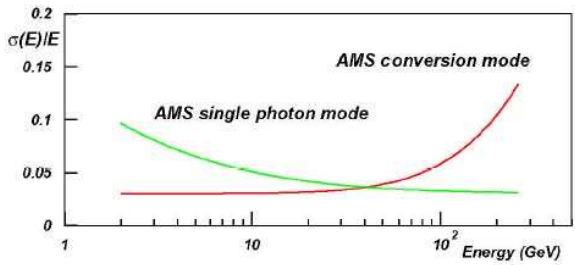}
\label{fig:Energy_res}}
\hfil
\subfigure[ Angular resolution]{\includegraphics[width=0.44\textwidth]{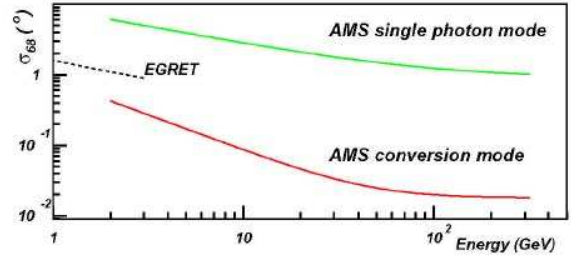}
\label{fig:Angular_res}}}
\caption{ The two modes of gamma detection in AMS-02.}
\label{fig:resol}
\end{figure*}

Both modes are characterized by different acceptance, effective area,  energy and angular resolutions,
as shown in Fig. \ref{fig:resol}. Their acceptances are comparable: $\rm 0.05-0.06~m^2sr$.

\begin{figure*}
\centerline{\subfigure[One year tracker sensitivity]{\includegraphics[width=0.44\textwidth]{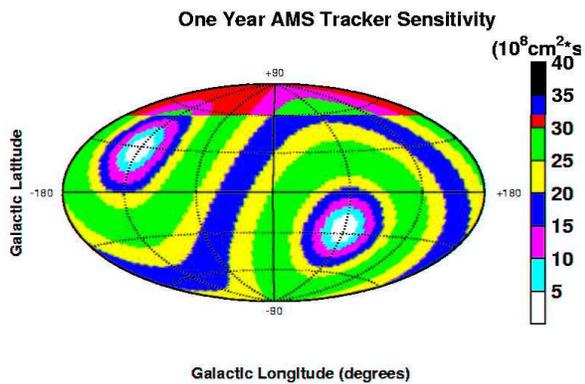}
\label{fig:Energy_res}}
\hfil
\subfigure[ One year Ecal sensitivity]{\includegraphics[width=0.44\textwidth]{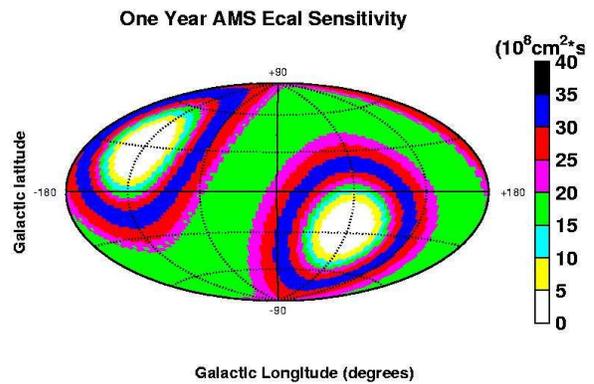}
\label{fig:Angular_re}}}
\caption{ Maps of sensitivity of the AMS-02 detector
for the two modes of gamma detection.
The time intervals when ISS orbits over South Atlantic Anomaly region are subtracted.}
\label{fig2}
\end{figure*}

The AMS-02 orbit and inclination angle are determined by the fact that the detector is
rigidly attached to the ISS.
From orbit simulation \cite{Nacho-orbit} the total exposure time of the detector
for the different sky regions is calculated.
Convolution of observation time (${\rm T_{obs}}$) and acceptance (${\rm A(E,d\phi) \times T_{obs}}$) gives
the sensitivity of AMS-02 for the considered regions of the sky.
Maps of the sensitivity are presented in Fig. \ref{fig2}.

The main background for gamma detection in both detection modes are protons,
due to their high abundance in cosmic rays.
Simulations show that the background rejection factor at the level of
${\rm O(10^6)}$ can be achieved. This will allow to keep background-to-signal ratio at the level of a few percent.

\par AMS-02 will have a star tracker on board. This device will allow to determine the reference system for
the incoming photons with accuracy better than a few arcminutes. This, together with
a very good angular resolution, will allow to localise sources with accuracy
better than 2 arcminutes.

\begin{figure}[h]
\begin{center}
\includegraphics*[width=0.44\textwidth,height=0.30\textheight,angle=0,clip]{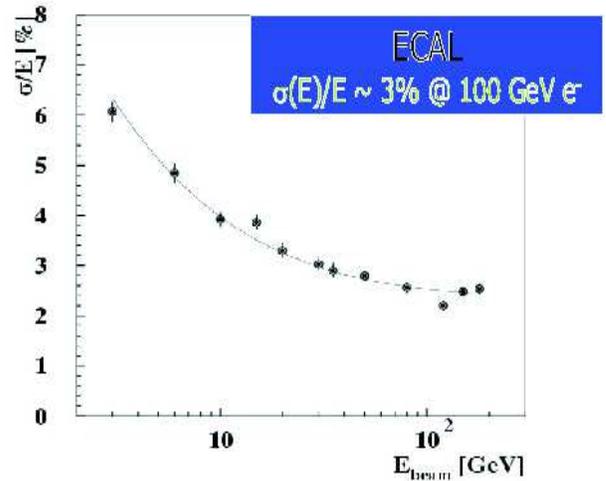}
\vspace{4mm}
\caption{\label {Test_cal} The measured energy resolution  as function of electron energy. }
\end{center}
\end{figure}

\begin{figure*}
\centerline{\subfigure[ Energy resolution]{\includegraphics[width=0.44\textwidth]{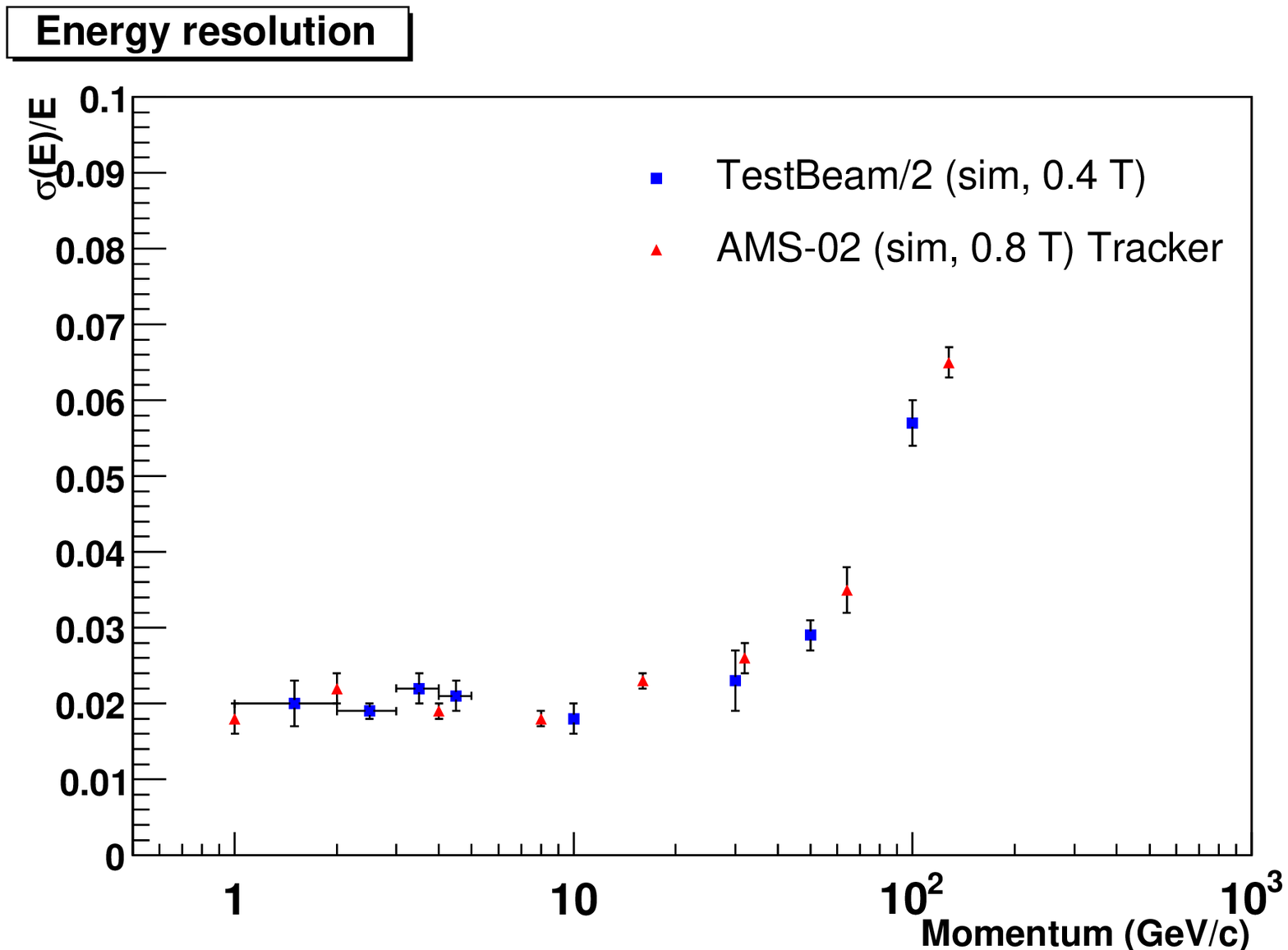}
\label{fig:Energy_res_beam}}
\hfil
\subfigure[ Angular resolution]{\includegraphics[width=0.44\textwidth]{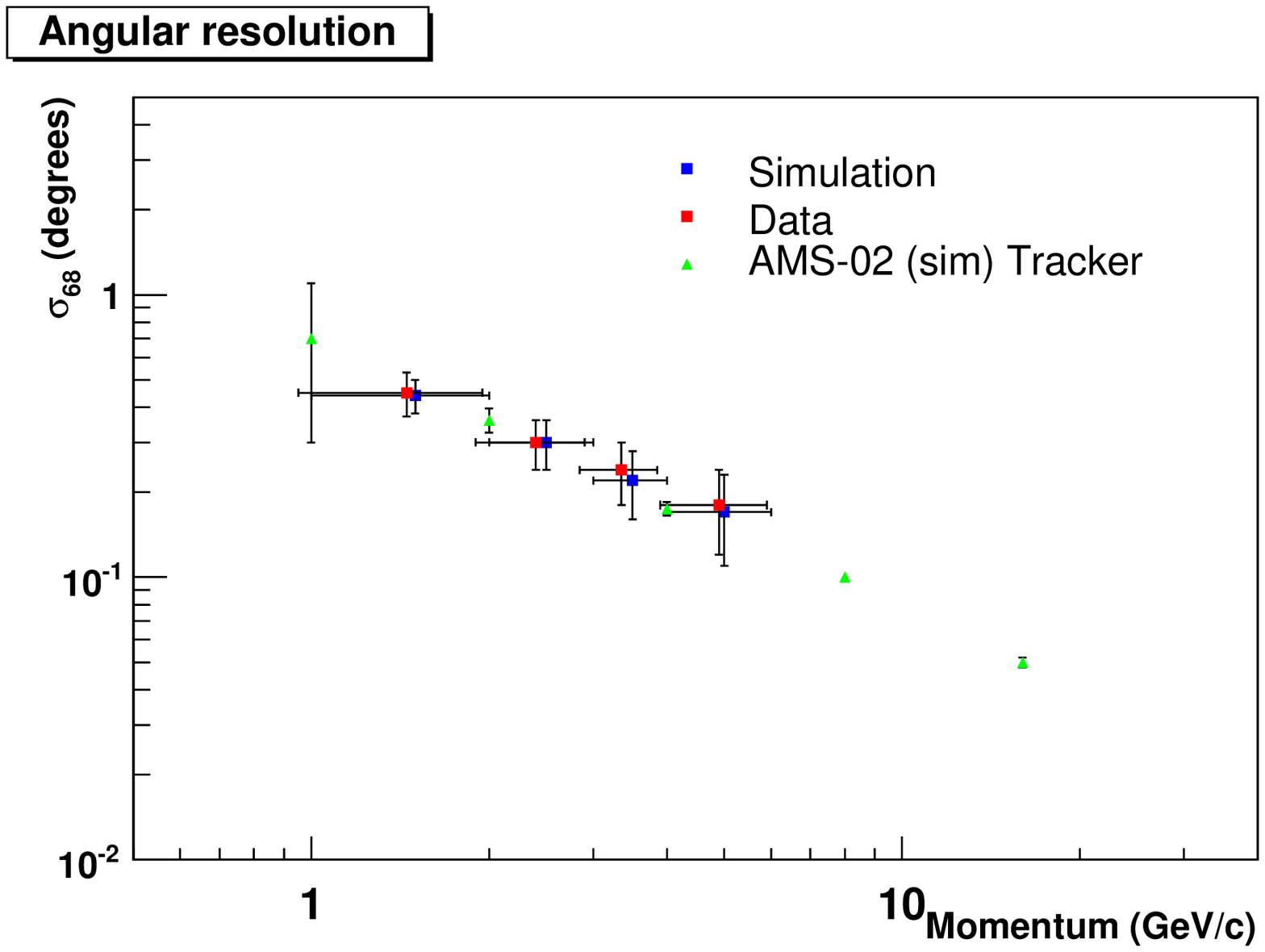}
\label{fig:Angular_res_beam}}}
\caption{ Energy and angular resolution measured at test beam for {\it conversion mode} compared with Monte Carlo expectations with 0.4 T magnetic field and extrapolated to field value of AMS-02 experiment, 0.8 T.}
\label{fig:test_beam}
\end{figure*}

The above detector performance were obtained from detector simulation
with use of GEANT \cite{GEANT} simulation software and tested in numerous beam tests.
The energy resolution of electron and  $\gamma$ detected in {\it Calorimetric mode} is measured with an electron test beam.
The results are shown in Fig.\ref{Test_cal}.
 The angular and energy resolution  
in conversion mode has been recently tested  with $\gamma$ produced from an electron beam \cite{Nacho-res} as shown 
in  Fig. \ref{fig:test_beam}.
%
%\begin{figure}[h]
%\begin{center}
%\includegraphics*[width=0.44\textwidth,height=0.30\textheight,angle=0,clip]{fig/eneres_tbeam.eps}
%\includegraphics*[width=0.44\textwidth,height=0.30\textheight,angle=0,clip]{fig/angres_tbeam.eps}
%\vspace{4mm}
%\caption{\label {test_beam}  Energy and angular resolution measured at test beam for {\it Conversion mode} }
%\end{center}
%\end{figure}

%There exists a software package (AMS Fast Simulator \cite{AMSFS}) for fast estimation of 
%the \hbox{AMS-02} potential in detection of various gamma ray phenomena. 
%It includes a parametrization of the detector response.
Many studies presented in the next session are performed using the AMS Fast Simulator \cite{AMSFS}.
It includes a parametrisation of the detector response suitable for a fast estimation of the potential in gamma rays detection.

%%%%%%%%%%%%%%%%%%%%%%%%%%%%%%%%%%%%%%%%%%%%%%%%%%%%%%%%%%%%
\section{Astrophysical gamma sources}
The 3rd EGRET catalogue \cite{3eg} contains 271 sources among which 170 are unidentified. 
With larger statistics and a different energy range AMS-02 will provide interesting data
about unidentified sources as well as known pulsars, blazars and AGNs. 
The very good angular resolution in {\it conversion mode}
of AMS-02 will allow to localise sources more precisely and determine their flux
with better accuracy due to better separation from diffuse background.
Discovery of new sources is also expected.

\begin{figure*}
\centerline{\subfigure[ Vela pulsar]{\includegraphics[width=0.44\textwidth]{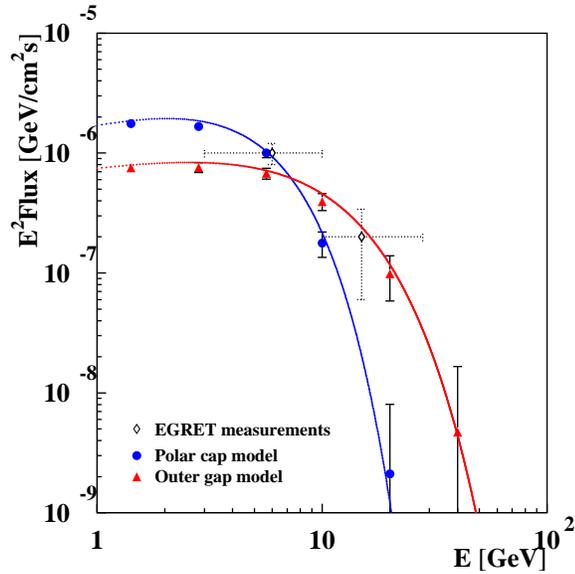}
\label{fig:Velamods}}
\hfil
\subfigure[ Diffuse gamma emission from the central part of the Galaxy ]{\includegraphics[width=0.44\textwidth]{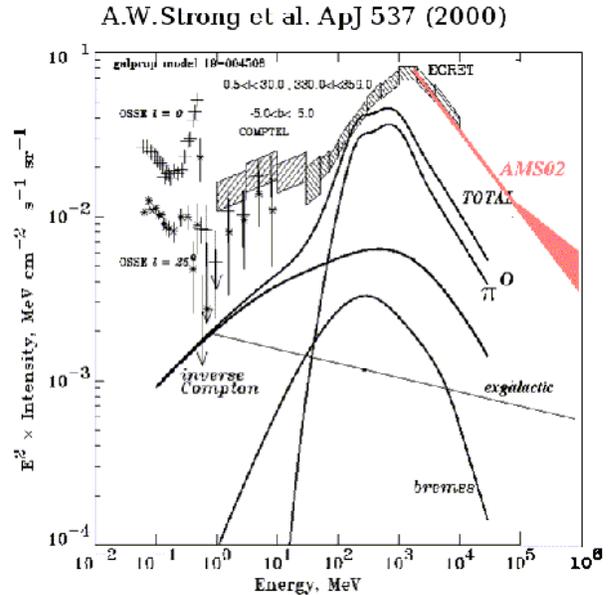}
\label{fig:Diffusion}}}
\caption{Left plot: expectations from {\it Outer Gap} and {\it Polar Cap} models of 
gamma ray emission from the Vela pulsar. 
AMS-02 will be able to distinguish between these two models \cite{Velamods}. 
Right plot: measurements and model predictions for diffuse gamma emission from the central part of the
Galaxy \cite{Strong:1998fr}. Estimated AMS-02 measurements are shown in red. }
\label{fig:model}
\end{figure*}

For example a possible AMS-02 measurement is presented on the left plot of Figure \ref{fig:Velamods}.
AMS-02 data on the spectrum of Vela pulsar  in the  energy range from 5 to 50 GeV, where there is
not enough statistics from EGRET, will allow to distinguish between two
models of gamma emission ~\cite{Velamods}.
%%%%%%%%%%%%%%%%%%%%%%%%%%%%%%%%%%%%%%%%%%%%%%%%%%%%%%%%%%%%%%%%%%%%%%%%%%%%%%%%%%%%%%%%%%%%%%

%\begin{figure}[h]
%\begin{center}
%\includegraphics*[width=0.44\textwidth,height=0.30\textheight,angle=0,clip]{fig/sp_psr_vela_1an.eps}
%\includegraphics*[width=0.44\textwidth,height=0.30\textheight,angle=0,clip]{fig/gamma.eps}
%\vspace{4mm}
%\caption{\label {fig3} Left plot: expectations from {\it Outer Gap} and {\it Polar Cap} models of 
%gamma ray emission from the Vela pulsar. 
%AMS-02 will be able to distinguish between these two models \cite{Velamods}. 
%Right plot: measurements and model predictions for diffuse gamma emission from the central part of the
%Galaxy \cite{Strong:1998fr}. Estimated AMS-02 measurements are shown in red.}
%\end{center}
%\end{figure}

%%%%%%%%%%%%%%%%%%%%%%%%%%%%%%%%%%%%%%%%%%%%%%%%%%%%%%%%%%%%
\section{Diffuse gamma background}
The galactic gamma ray diffuse background is believed to be produced in the interstellar medium
by $\pi^0$ decay, bremsstrahlung and inverse Compton scattering.  
It has been measured by EGRET \cite{Hunger:1997we}
and is usually presented as integrated flux in energy bins. 
It allows not only to create sky maps of the diffuse background but also
to perform spectral studies. On the right plot of Figure \ref{fig:Diffusion} 
the gamma ray spectrum observed by EGRET and other experiments 
is shown together with a gamma ray emission model. 
The EGRET measurements present a flux excess in the GeV energy region. 
This is explained in various ways, from model tuning \cite{Strong:2004ry} to
assumption of Dark Matter annihilation \cite{deBoer:2004es}.
\hbox{AMS-02} will contribute to a better understanding of the diffuse gamma
spectrum in the GeV region.

The improvement of this measurement in the GeV-region could
lead to conclusion about the nature of this excess.

The maps of the gamma ray background which will be measured by AMS-02, 
are expected to have better resolution than the EGRET ones. 
It will help in observations and flux determination of the gamma sources.

%%%%%%%%%%%%%%%%%%%%%%%%%%%%%%%%%%%%%%%%%%%%%%%%%%%%%%%%%%%%
\section{Dark Matter}
The nature of Dark Matter is one of the outstanding questions of cosmology. 
Its existence is required by a multitude of observations~\cite{Bahcall:1998ur}. 
In the framework of the Standard Cosmological Model, WMAP \cite{Bennett:2003bz} quotes a total 
matter  density  $\Omega_m = 0.27 \pm 0.04$ and baryon density $\Omega_b = 0.044 \pm 0.04$, 
which confirm that most of the matter is non-baryonic, in agreement with the results obtained 
from primordial nucleosynthesis.  
Supersymmetric theories offer an excellent Weakly Interacting Massive Particle (WIMP) as a Dark Matter component.
 In particularly the neutralino ($\chi^0_1$) of the Minimal SuperSymmetric Standard Model (MSSM) assumed as  Lightest Supersymmetric Particle (LSP) is 
the most accredited candidate. Less conventional scenarios than the neutralino in the minimal supersymmetric model (mSUGRA) as  Anomaly Mediated Supersymmetry Breaking (AMSB)  and extra-dimensional models as Kaluza-Klein particles have been also suggested (Ref. \cite{Jacholkowska:2005nz} and references therein).

AMS-02 has potential in observation of Dark Matter signal from the Galactic Center in the gamma channel.
The expected signature is a deformation of the spectrum or a monochromatic line 
from $\chi \chi \rightarrow \gamma \gamma, Z\gamma$ processes.

\par The $\gamma$-predicted fluxes from Galactic Centre from neutralino annihilation in the framework of mSUGRA, 
AMSB and Kaluza-Klein  dark matter annihilation  for different scenarios are used to define the discovery potential 
for non-baronic Dark Matter by \hbox{AMS-02} \cite{Jacholkowska:2005nz}. 
This  probe of the Dark Matter trough indirect $\gamma$ detection  depends on astrophysical uncertainty
%the density profile near Galactic Center 
and  
%from the assumptions 
on the particle physics models.
\par
\begin{figure*}
\centerline{\subfigure[NFW profile]{\includegraphics[width=0.44\textwidth]{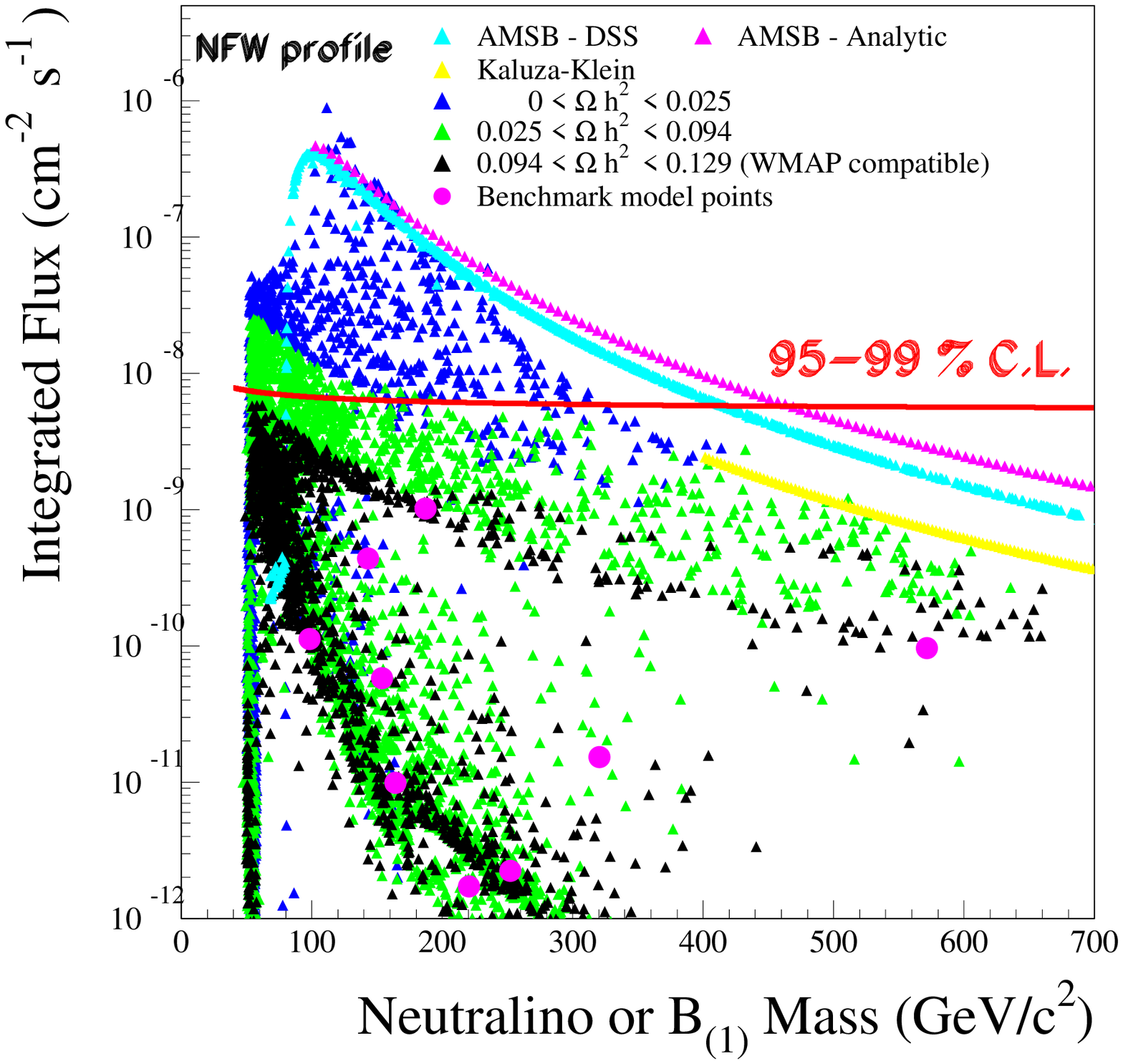}
\label{fig:NFW}}
\hfil
\subfigure[NFW cuspy profile]{\includegraphics[width=0.44\textwidth]{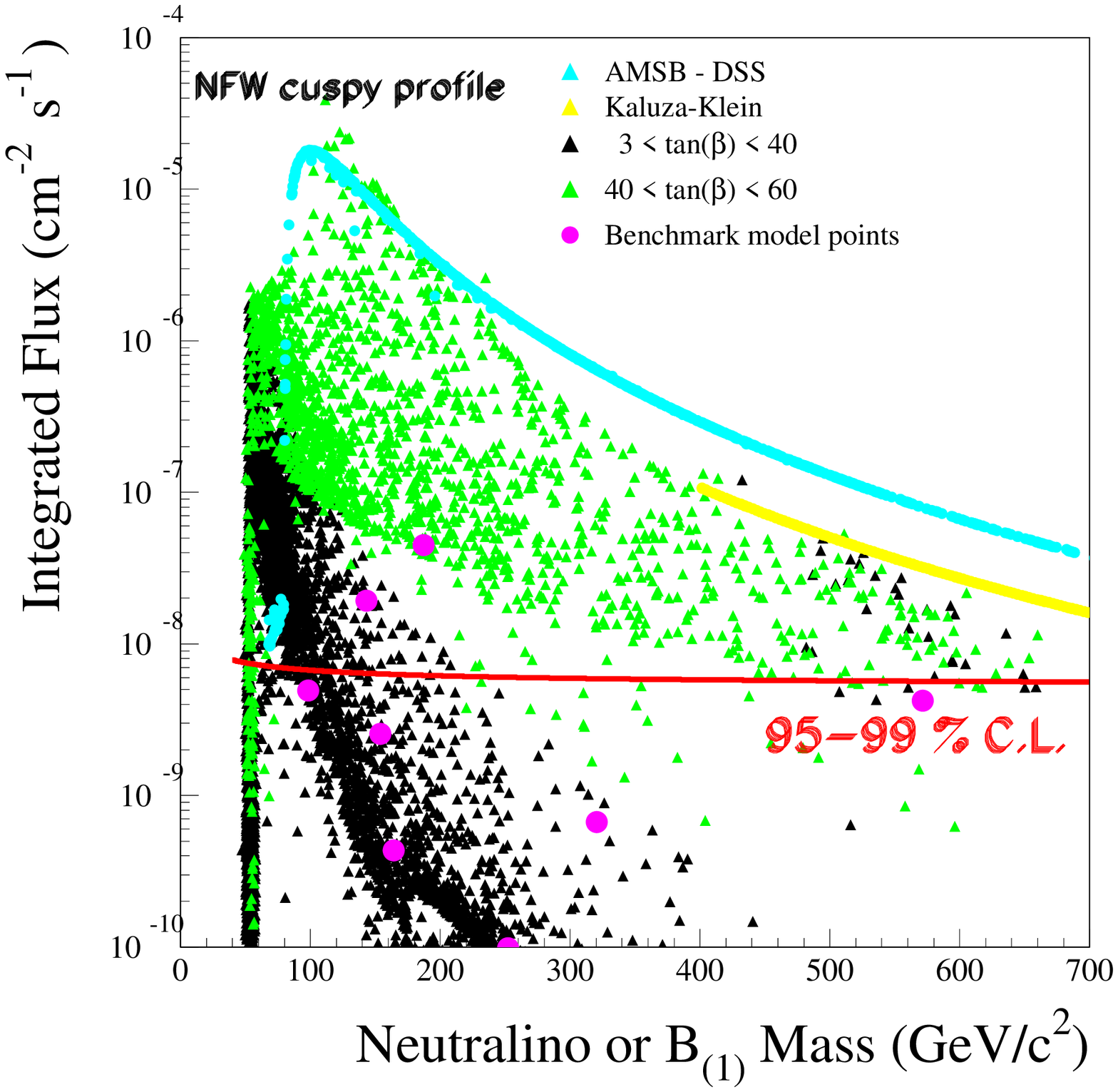}
\label{fig:NFW_cuspy}}}
\caption{ The integrated $\gamma$ flux from Galactic Centre as function of $\rm {m_{\chi}}$ for NFW(right) and NFW cuspy (left) halo profile parameterisations ~\cite {Jacholkowska:2005nz}.}
\label{fig:NFW_total}
\end{figure*}

 The particle  physics models can be grouped as:
\begin{itemize}
\item {} {\bf mSugra models}
 In this framework, few  SUSY benchmark points have been  proposed to provided a common way of comparing 
 the SUSY discovery potential of accelerators. 
 Each of these points corresponds to different configuration of five mSUGRA parameters \cite{Jacholkowska:2005nz}.   
 The integrated $\gamma$ flux from the Galactic Centre as function of neutralino mass $\rm{M_\chi}$ 
 for a NFW standard profile and for a $\gamma$-ray energy threshold of 1 GeV is shown in 
 Fig~\ref {fig:NFW}. 
 Fig.~\ref {fig:NFW_cuspy} shows the results for a more favorable NFW cuspy dark matter profile.
\item {} {\bf Anomaly Mediated Supersymmetry Breaking parametrization},
as the name suggests, is a gravity-mediated mechanism. 
In these models the lightest neutralino and the lightest chargino are mass-degenerate.
\item {} {\bf Kaluza-Klein Dark Matter}. 
These models with compact extra-dimensions predict several new states called Kaluza-Klein (KK) excitations.
The conservation of KK number implies that the particles of this model cannot decay in Standard Model particles, 
thus the lightest particle is stable. 
The most promising  lightest particle candidate for Dark Matter is the first excitation level of Hypercharge 
gauge boson, $\rm B_{(1)}$.

\end{itemize}

\par The  astrophysical uncertainty depends on the assumptions  of density profile near Galactic Center.
% The $gamma$-ray fluxes depend on the assumptions  of density profile near Galactic Center.
The mass density profile of the Galaxy, $\rho_\chi(r) $, can be parametrized assuming a simple spherical Galactic 
halo as function of  the distance of Galactic Centre from Earth, $R_0$,the halo density, $\rho_0$, 
and the core radius $a$. The value of the three parameter $\alpha$, $\epsilon$ and $\gamma$  distinguish the different models.
\begin{equation}
% \rho_\chi(r)  = \rho_0 \Bigl( \frac{R_0}{r} \Bigr) 
\rho_\chi(r)  = \rho_0 \left( \frac{R_0}{r} \right)^\gamma 
\left({\frac {R_0+ a^\alpha}{r^\alpha + a^\alpha}} \right)^\epsilon
\end{equation}
The NFW-standard, NFW-cuspy and Moore profile models differ among them for the value of these parameters. 
The increased number of photons corresponds to NFW-cuspy profile which is the most favorable to this study.

The expected fluxes for different modes together with AMS sensitivity 
 are shown in Fig.  ~\ref {fig:NFW_total} for  different models.
 More on this subject can be found in \cite{John-DM, Jacholkowska:2005nz}.

%%%%%%%%%%%%%%%%%%%%%%%%%%%%%%%%%%%%%%%%%%%%%%%%%%%%%%%%%%%%
\section{Gamma Ray Bursts}

Gamma Ray Bursts are the most energetic phenomena in the Universe. 
Their nature, after many years of observations 
and collection of rich GRB catalogs, remains unexplained.
More observations are needed 
especially at high energy as the dynamic range 
of previous experiments extends only to around 1 GeV.

Due to the large AMS-02 acceptance and to the relatively large field of view, 
observation of a few GRBs during the 3 year mission is expected. 
AMS-02 observations lie in the unexplored energy range.
An extrapolation from previous GRBs allows to estimate that about
30 gammas above 1 GeV will be measured if a burst similar to GRB950503 happens again.
In addition, at the time of AMS-02 mission, other dedicated GRB observatories will be 
on orbit. A synchronization of observations (one of the main reasons for the presence of a dedicated 
Global Positioning System on board AMS-02) might lead to interesting results. 
The good time resolution of the detector will also allow to perform some studies
on quantum structure of the space-time \cite{Amelino-Camelia:1997gz}.

%%%%%%%%%%%%%%%%%%%%%%%%%%%%%%%%%%%%%%%%%%%%%%%%%%%%%%%%%%%%%%
\section{Conclusion}
The AMS-02 gamma ray physics program is rich. One of the most important issue is Dark Matter search.
The  Dark Matter indirect search potential by AMS experiment is explored with an evaluation of
$\gamma$-ray flux  from Galactic Centre Region, in some benchmark points
of mSUGRA,  AMSB  and Kaluza-Klein models. Only models with mSUGRA scenario in particular the
region of supersymmetric parameter space yield significant results on a realistic time scale.
Several mSUGRA models in the case of a favorable galactic halo configuration, such as the cuspy and very cuspy NFW
profiles, provide exclusion limits after three years of data taking.

The AMS-02 gamma ray physics program  includes also  the mapping of the gamma ray diffuse background,
observation of gamma ray sources and measurements of their spectra, and of rare, high energy Gamma Ray Bursts.
In all this domain the AMS contribution will be fundamental and these are  little steps in the Universe understanding.

\section*{Acknowledgment}

I would like to thank for the discussions in the preparation
of this contribution and for the critical
reading of manuscript Dr. Mariusz Sapinski.

\end{document}